# Phase transition in CaFeAsH:

# bridging 1111 and 122 iron-based superconductors

*Yoshinori Muraba[†], Soshi Iimura[‡] Satoru Matsuishi[†], Hidenori Hiramatsu[†, ‡]*

*Takashi Honda[§,¶], Kazutaka Ikeda[§,¶], Toshiya Otomo,[§ ,¶]Hideo Hosono\*[†, ‡]*

[†]Materials Research Center for Element Strategy, Tokyo Institute of Technology, 4259 Nagatsuta-cho, Midori-ku, Yokohama 226-8503, Japan

[‡]Materials and Structures Laboratory, Tokyo Institute of Technology, 4259 Nagatsuta-cho, Midori-ku, Yokohama 226-8503, Japan

[§] Institute of Materials Structure Science, High Energy Accelerator Research Organization (KEK), Tsukuba, Ibaraki 305-0801, Japan

[¶] J-PARC Center, KEK, Tokai, 319-1106, Japan

Corresponding author email: hosono@msl.titech.ac.jp

Abstract

Iron-based superconductors can be categorized as two types of parent compounds by considering the nature of their temperature-induced phase transitions; namely, first order transitions for 122- and 11-type compounds and second-order transitions for 1111-type compounds. This work examines the structural and magnetic transitions (ST and MT) of CaFeAsH by specific heat, X-ray diffraction, neutron diffraction, and electrical resistivity measurements. Heat capacity measurements revealed a second-order phase transition accompanies an apparent single peak at 96 K. However, a clear ST from the tetragonal to orthorhombic phase and a MT from the paramagnetic to antiferromagnetic phase were detected. The structural ($T_s$) and Néel temperatures ($T_N$) were respectively determined to be 95(2) and 96 K by X-ray and neutron diffraction and resistivity measurements. This small temperature difference, $T_s - T_N$, was attributed to strong magnetic coupling in the inter-layer direction owing to CaFeAsH having the shortest lattice constant $c$ among parent 1111-type iron arsenides. Considering that a first-order transition takes place in 11- and 122-type compounds with a short inter-layer distance, we conclude that the nature of the ST and MT in CaFeAsH is intermediate in character, between the second-order transition for 1111-type compounds and the first-order transition for other 11- and 122-type compounds.

I. Introduction

Current interest in iron-based superconductors is focused on developing an understanding of the mechanism of superconductivity and determining the relationship between structural transitions (STs) and magnetic transitions (MTs) of the so-called parent compounds.[1–3] Typical iron-based superconductors compounds include 122-type $Ae$Fe$_2$As$_2$ ($Ae$ = alkaline earth and Eu) [4–9] and 1111-type $Ln$FeAsO ($Ln$ = lanthanide)[10–18] or $Ae$FeAs$X$ ($X$ = H, F)[19–23]. Their crystal structures consist of an alternating stack of an [FeAs]$^-$ conducting layer and a blocking layer such as $Ae^{2+}$, [$Ln$O]$^+$, or [$AeX$]$^+$. For the parent 122-type compounds, a first-order ST from a tetragonal to orthorhombic phase and a MT from a paramagnetic to antiferromagnetic (AFM) take place at the same transition temperature above 100 K.[24–27] However, for 1111-type compounds, these transitions are of a second-order-type and the transition temperatures are distinctly separated [i.e., the ST ($T_s$) is greater than the Neel temperature ($T_N$)].[28,29] Superconductivity emerges when two transitions are suppressed by carrier doping or applied pressure.

The ST and MT transitions of 122-type compounds may be classified into three regions depending on the carrier concentration and the applied pressure;[30,31] region (i) in which the first-order ST and MT occur simultaneously in the parent compounds. Region (ii) for which the second-order ST occurs at a $T_s$ preceding the first-order MT ($T_s > T_N$) owing to light carrier doping or applied pressure. Region (iii) for which the difference between $T_s$ and $T_N$ is expanded and the MT becomes second-order through a further increase of doping or pressure. Contrary to the 122-type compounds, the ST and MT of 1111-type compounds remain in region (iii) irrespective of doping or pressure until both transitions are suppressed. Theoretically, this rich evolution of ST and MT with doping, pressure, and FeAs interlayer distance is well described by a magnetic scenario based on an itinerant spin model. In this model, STs and MTs stem from magnetic interactions, and the characteristics of the ST and MT change from the first-order [region (i)] to second-order [region (iii)] as the shear modulus ($C_s$) and dimensionality ($d$) respectively become larger and smaller. [1,2,32,33]

In this paper, we study ST and MT of CaFeAsH, which possesses the following two features. The first is the smallest bulk modulus, 61.3GPa, among 1111-type compounds, which is comparable to that of $BaFe_2As_2$ (57.7GPa).[34–38] Because the bulk modulus of isostructural CaFeAsF (82.8 GPa) is larger than that of CaFeAsH, the softness of CaFeAsH is attributed to compressible hydride ions ($H^-$) incorporated into the blocking layer. The second feature is the short interlayer distance, which is among the shortest of 1111-type compounds. As a consequence, a 3-dimensional Fermi surface, similar to those of the 122-type has been confirmed by DFT calculations.[22,39,40] These features of CaFeAsH suggest that the characteristics of the ST and MT will be closer to those of 122-

type compounds rather than those of 1111-type compounds.

We investigated $T_s$ and $T_N$ of the parent CaFeAsH. First, we show that CaFeAsH has only a single second-order transition at approximately 96 K, as revealed by specific heat measurements. We verify that the $T_s$ of CaFeAsH is located between 93 and 97 K based on low-temperature X-ray diffraction measurements. The $T_N$ is determined to be 96 K by electrical resistivity measurements and observations of stripe-type AFM ordering, which is commonly observed in 122- and 1111-type compounds, as confirmed by neutron diffraction performed at 10 K. The ST and MT characteristics of CaFeAsH are discussed in terms of structural and local magnetic aspects by comparison with those features of other 1111-type and 11-, 111- and 122-type compounds. Finally, we conclude that the small temperature difference, $T_s - T_N$, of CaFeAsH can be attributed to the strong magnetic coupling along the inter-layer direction derived from it having the shortest lattice constant $c$ among parent 1111-type iron arsenides.

II. Experimental

CaFeAsH or CaFeAsD were prepared from CaAs, CaH$_2$ (or CaD$_2$), and Fe$_2$As,

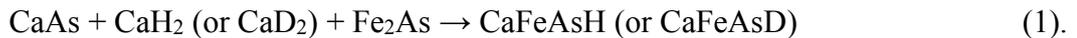
CaAs + CaH$_2$ (or CaD$_2$) + Fe$_2$As → CaFeAsH (or CaFeAsD)     (1).

CaAs and Fe$_2$As were synthesized from their respective metals, and CaH$_2$ (or CaD$_2$) was obtained by heating metal Ca in a H$_2$ (or D$_2$) atmosphere. All the starting materials and precursors were prepared in a glove box under a purified Ar atmosphere (dew point < − 90 °C and oxygen concentration < 1 ppm). A pellet of the ground starting materials was placed in a boron nitride capsule, sandwiched by two pellets of Ca(OH)$_2$ [or Ca(OD)$_2$] and a NaBH$_4$ (or NaBD$_4$) mixture acted as an excess solid hydrogen source.[41] The final product was synthesized at 1173 K under 2.5 GPa for 30 min with the use of a belt-type high-pressure anvil cell.[42]

The temperature dependence of the specific heat ($C_p$) was measured in the temperature range of 2 to 300 K by a relaxation method with the use of a physical property measurement system (PPMS, Quantum Design, Ltd.). The temperature rise for each relaxation was set to be 2% of the sample temperature. The relaxation method estimates the specific heat of samples by measuring the thermal response and analyzing its relaxation process.[43,44] An advantage of this method is that it requires a smaller amount of sample (a few milligram) than the adiabatic method which requires a few gram of sample and the experienced technique.[45] In spite of the simplicity of measurement, the measured heat capacity by relaxation technique falls within a range of ± 2% deviation of values measured by adiabatic technique which provides the most accurate data among the all specific heat measurements. On the other hand, when the relaxation is across a first order transition temperature, the estimated specific heat is significantly underestimated compared to the value measured by adiabatic technique. However, the disadvantage is resolved by adopting scanning analysis described in results section. The electrical resistivity ($\rho$) measurements were performed in the same temperature range with the PPMS. XRD patterns were measured for CaFeAsH sealed in a glass capillary with the diameter of 0.5 mm using a Bruker D8 Advance with monochromatized Mo K$\alpha_1$ radiation ($\lambda$ = 0.7093 Å). For low temperature measurements, the capillaries were cooled by gas-flow methods using cold $N_2$ gas above 95 K and He gas below 95 K. Neutron powder diffraction (NPD) was performed with the neutron total scattering spectrometer (NOVA) installed in Materials and Life Science Experimental Facility (MLF) of the Japan Proton Accelerator Research Complex (J-PARC). A deuterium-substituted sample, CaFeAsD, was used to avoid the high background derived from H, which shows considerable incoherent scattering. The sample was loaded into a vanadium-nickel alloy holder (6.0 mm inner diameter and 0.1 mm wall thickness) sealed by an indium wire. Powder neutron diffraction data were collected at 150 and 10 K. The XRD and NPD data were analyzed by the Rietveld method with the TOPAS and Fullprof codes, respectively.[46,47] Crystal and magnetic structures were visualized with the use of the computer program VESTA.[48]

III. Results

Specific heat measurements are a direct method of determining the phase transition temperature. **Figure 1(a)** shows the temperature dependence of the $C_\mathrm{p}$ of CaFeAsH. Unlike the double-peak-structure of conventional 1111-type compounds,[28,29,49–51] only a single peak was observed at approximately 96 K, indicating that the two transition temperatures of ST and MT are located close to each other or merged. To estimate the transition entropy, the lattice and electronic contributions were extracted from the data over the whole temperature range. We estimated these contributions using the function given in Eq. (2),

$$C_\mathrm{p}(T) = \gamma T + k\, C_\mathrm{D}(T, \Theta_\mathrm{D}) + (1 - k)C_\mathrm{E}(T, \Theta_\mathrm{E}), \qquad (2)$$

where $\gamma$ is the Sommerfeld coefficient, $C_\mathrm{D}$ and $C_\mathrm{E}$ are the Debye and Einstein terms, respectively, and $k$ determines the contribution of the Debye term to specific heat. The parameters $C_\mathrm{D}$ and $C_\mathrm{E}$ are respectively given by Eq. (3) and (4):

$$C_D(T, \Theta_D) = 9NR \left(\frac{T}{\Theta_D}\right)^3 \int_0^{\frac{\Theta_D}{T}} \frac{x^4\, exp(x)}{\{exp(x)-1\}^2}\, dx\,, \qquad (3)$$

$$C_E(T, \Theta_E) = 3NR \left(\frac{\Theta_E}{T}\right)^2 exp\left(\frac{\Theta_E}{T}\right) \left\{exp\left(\frac{\Theta_E}{T}\right) - 1\right\}^{-2}, \qquad (4)$$

where $N$ is the number of atoms per formula unit ($N = 4$ for orthorhombic CaFeAsH) and $R$ is the molar gas constant. Generally, in order to reproduce the experimental specific heat at high temperature region, the contribution of excited optical modes have to be included as well as the acoustical modes which are expressed by the Debye model.[52] These optical modes are usually approximated by the Einstein model. For various iron arsenides, the linear combination of Debye and Einstein terms successfully reproduces the specific heat, suggesting that the optical modes located at low frequency significantly affect the specific heat of iron arsenides measured for a temperature range from 2 to 300K. (for LaFeAsO, there are three optical modes at 100, 180 and 290 cm-1)[53–57] The parameters extracted from this fit were $\gamma$ = 8.18(19) mJ/molK, $\Theta_D$ = 324 (1) K, $\Theta_E$ = 823 (6) K and $k$ = 0.716 (2). The $C_p$ change ($\Delta C_p$) was calculated by subtracting the fitting curve from the $C_p$ data. The entropy change ($\Delta S$) through the phase transition was determined by integration of $\Delta C_p/T$ [**Figure 1 (b)**]. The obtained value of $\Delta S$ gradually increased on heating and reached ~ 0.6 J/molK, which is slightly larger than the values associated with STs and MTs observed in parent 1111-type compounds (e.g., 0.27 and 0.5 J/molK for LaFeAsO and SrFeAsF, respectively),[28,58] and smaller than those of conventional 122- and 11-type compounds (e.g., 1.0 and 3.2 J/molK for SrFe$_2$As$_2$ and FeTe, respectively).[25,59]

In the normal relaxation method, the specific heat at each temperature is obtained by fitting the temperature-time relaxation curve. However, if the sample temperature crosses the first-order phase transition point, the temperature relaxation curve is deformed, as illustrated in **Figure 1(c)**, and the specific heat cannot be accurately determined by this relaxation method. In **Figure 1(d)**, we plot the observed temperature-time relaxation curve across the specific heat jump. There is no plateau region derived from the latent heat associated with the first-order phase transition[24–26,59] and the temperature-time relaxation curve is well fitted by a linear combination of two exponential functions with different time constants.

A scanning method is an alternative way to estimate the specific heat, especially close to the first-order phase transition.[60] Because the scanning method directly obtains the specific heat from the time differential of the relaxation data, it can correctly trace the steep change of the specific heat at the first-order transition; however, this parameter is always underestimated by the normal relaxation method. In **Figure 1 (e)** we compare the two $C_p$ ($T$) curves calculated by the two methods. The coincidence of the two specific heat curves and the gradual $\varDelta S$ increase at the transition temperature, as shown in **Figure 1 (b)**, indicate that the ST and MT of CaFeAsH are second-order.

**Figure 2(a)** shows the temperature variation of the XRD peak profiles, having a tetragonal 221 reflection for CaFeAsH. In many cases, a split in the tetragonal 220 reflection peak is observed on detection of the ST for 1111- and 122-type compounds. We examined the tetragonal 221 reflection in our measurements; however, because the positions of the 220 and 006 reflections closely overlapped each other a clear splitting of the 220 reflection could not be observed. Instead, the 221 reflection clearly split into the orthorhombic 401 and 041 reflections in the temperature range of 95 ≤ $T$ < 100 K. To determine $T_s$ of CaFeAsH, we first fitted the XRD patterns to an orthorhombic *Cmme* structure and then calculated an order parameter, $P = (a_O+b_O)/(a_O–b_O)\times 10^3$, where $a_O$ and $b_O$ are lattice constants of *Cmme* structure [**Figure 2 (b)**]. The order parameter increased sharply at 95 K on cooling, and approached 3 at 70 K, which is a value comparable to those of 1111-type *Ln*FeAsO (2.4−2.8) and *Ae*FeAsF (~3.25−3.4).[61–69] We fitted the data to an empirical power low function $f(T) = \alpha(1 − T/T_s)^\beta$ and obtained $\beta = 0.1$ (2) and $T_s =$ 95 (2) K, which was consistent with the $C_p$ peak at 96 K. **Figure 2(c)** shows the temperature dependence of the lattice parameters of CaFeAsH. We plotted the tetragonal lattice constants for $T > T_s$ and the orthorhombic lattice constants for $T < T_s$. In addition to the *ab* splitting below $T_s$, a small decrease of the *c*-axis was also detected.

Next, we discuss the magnetism of CaFeAsH. **Figures 3(a)** and **(b)** show NPD patterns for a high *d*-spacing region at 150 and 10 K, respectively. At 10 K, extra peaks appeared at $d = 3.9$ and 5.2 Å. These magnetic peaks were indexed by conventional stripe-type AFM ordering to an orthorhombic *Cmme* lattice, as shown in **Figure 3(d)**.[70,71] The spins on the Fe atoms were aligned along $a_O$ axis (> $b_O$ axis), and antiferromagnetically coupled along the $c_O$ axis. The propagation vector was determined to be (1, 0, ½) on the orthorhombic lattice, and the magnetic moment on Fe atom was refined to be 0.71 (1) $\mu_B$

/ Fe at 10 K.

The temperature dependence of the electrical resistivity $\rho$ for CaFeAsH was measured to determine $T_N$ of CaFeAsH (**Figure 4**). A sudden decrease was observed below 100 K, which is known to be intimately coupled to the STs and MTs of 1111- and 122- type iron pnictides. From previous research on the iron arsenides, it is known that a peak temperature of the $d\rho/dT - T$ curve corresponds to $T_N$.[28,29,49,50,72] Following this empirical rule, we estimated the $T_N$ of CaFeAsH to be 96 K. The small temperature difference between $T_s$ (= 95 (2) K) and $T_N$ (= 96 K) of CaFeAsH suggested that the single peak of $C_p$ ($T$) at 96 K was composed of ST and MT.

IV. Discussion

First, we discuss the small temperature difference of $T_s - T_N$ in CaFeAsH by comparing the isostructural CaFeAsF with the $T_s - T_N$ value of 20 K. According to the magnetic scenario, the shear modulus ($C_s$) and the dimensionality ($d$) control the nature and the transition temperature of the ST and MT. Structural parameters and several physical properties for CaFeAsH and CaFeAsF are summarized in **Table 1**. Unlike the deep F$^-$-2$p$ level, the shallow energy level of H$^-$-1$s$ energetically overlapped with the As$^{3-}$-4$p$ level to form a covalent bond between the [CaH]$^+$ and [FeAs]$^-$ layers, making the lattice constant $c$ of CaFeAsH decrease. In the two compounds, the lattice constant $c$ was the sum of the As height from the Fe plane ($h_{As}$), the Ca height from the H/F planes ($h_{Ca}$), and the separation between the Ca- and As-planes ($d_{Ca-As}$). The reduced $h_{Ca}$ and $d_{Ca-As}$ values compared with those of CaFeAsF were a consequence of H−As bonding. The shortened lattice constant $c$ decreased the dimensionality of the crystal and electronic structures. In this case the magnetic scenario predicts that the value of $T_s - T_N$ should decrease, which is consistent with the observed small temperature difference of CaFeAsH.

The bulk modulus ($Bm$) is proportional to the shear modulus ($Cs$) for a polycrystalline system. The 1111-type compounds consist of alternate stacking of the hard ionic blocking layer and the soft covalent layer. The hard blocking layer primarily determines the $Bm$ because it is less compressible than the conducting FeAs layer. In fact, the $Bm$ of 1111-type compounds is higher than those of 11, 111, and 122-type compounds (see Figure S1). Hence, the smaller $T_s - T_N$ of CaFeAsH than that of CaFeAsF can also be understood in terms of the smaller $Bm$ owing to substitution of H⁻ for F⁻ in CaFeAsF.

Next, we discuss the effect of the interlayer distance and $Bm$ on the $T_s - T_N$ of other iron-based superconductors, including 11-, 111-, and 122-type parent compounds. We first plotted $T_s - T_N$ versus their $Bm$ (Figure S1). Although the overall trend obeyed the theoretical prediction (*i.e.*, $Bm$ increases from ~30 GPa for FeTe to more than 80 GPa for 1111-type compounds), $T_s - T_N$ appeared to increase and it was difficult to find a clear relationship among these parameters. However, we found a clear correlation between $T_s$ − $T_N$ and the inter-layer distance, as shown in **Figure 5**. The value of $T_s$ − $T_N$ monotonically increased as the interlayer distance increased, except for 111-type compounds. The deviation of 111-type NaFeAs from the general trend could be attributed to carrier doping produced by a partial Na deficiency.[76] Owing to CaFeAsH having the shortest interlayer distance among 1111-type compounds, the $T_s$ − $T_N$ was located between other parent 1111-type and 122-, 11-, and 111-type compounds.

Finally, we quantitatively compared the characteristics of the STs and MTs of CaFeAsH with those of 122-, 111-, and 11-type compounds. If we adopted the Heisenberg-type local magnetic exchange model, the temperature difference between $T_s$ and $T_N$ of the parent compounds is directly calculated from only $J_1$, $J_2$, and $J_z$ (in-plane nearest, next nearest, and out-of-plane magnetic couplings, respectively) without considering the lattice softness.[3,61,73] In this model, stripe-type AFM ordering is successfully predicted when $J_1$ is less than $2J_2$, and the nature of the STs and MTs changes to region (i) and then to (iii) in turn as $J_z/J_2$ decreases. For electron-doped BaFe$_2$As$_2$, NaFeAs, and FeTe, $J_z/J_2$ is directly determined by inelastic neutron scattering of those single crystals. For the parent compounds, the $J_z/J_2$ values are close to each other (0.13, 0.095, and 0.046 for BaFe$_2$As$_2$, NaFeAs, and FeTe, respectively).[74–76] However, the slight electron-doping of BaFe$_2$As$_2$ strongly depresses $J_z/J_2$ from 0.13 ($e^-$/Fe = 0) to 4.9×10$^{-3}$ ($e^-$/Fe = 0.015) and 5×10$^{-4}$ ($e^-$/Fe = 0.04).[77,78] The boundary of region (ii) and region (iii) is located at approximately 0.022 $e^-$/Fe.[30,31] For 1111-type compounds, the theoretical value of $J_z/J_2$ is available in Ref. 60 as calculated from $(T_s - T_N)/T_N$. For this calculation, we set $T_s$ = 97 K to be the upper limit of $T_s$ = 95 (2), because the ST always precedes the MT ($T_s \geq T_N$) within this theory. Although the $T_s - T_N$ of CaFeAsH was smallest among the parent 1111-type compounds, the $J_z/J_2$ (1.3×10$^{-3}$) was smaller than those of parent 122-, 111-, and 11-type compounds and close to that of 0.04 $e^-$/Fe doped BaFe$_2$As$_2$. These considerations indicate that the ST and MT of CaFeAsH is located at region (iii), and the small $J_z$ is attributed to the thicker [CaH]$^+$ blocking layer that those of the layers formed by $A$ and $Ae$ cations in 111- and 122-types compounds, respectively.

Summary

We investigated structural and AFM transition of CaFeAsH by specific heat, NPD, XRD, and electrical resistivity measurements. Unlike the double-peak-structure of the specific heat in other parent 1111-type compounds, only a single and second-order transition was observed at approximately 96 K. The $T_s$ and $T_N$ values were determined to be 95(2) and 96K by low temperature XRD and resistivity measurements, respectively, which coincided well with the peak temperature of the $C_p$ measurement. The magnetic structure of CaFeAsH was determined by NPD to have a conventional stripe-type order with a refined magnetic moment of 0.71 (1) $\mu_B$ on the Fe atom. We attributed the observed small temperature difference, $T_s - T_N$, to the strong magnetic coupling along the inter-layer direction derived from CaFeAsH having the shortest lattice constant $c$ among the parent 1111 iron arsenides. Finally, we concluded that the nature of the ST and MT for CaFeAsH is intermediate between those of the second-order transitions of 1111-type compounds and the first-order transitions for other 11- and 122-type compounds. This understanding of the structural and magnetic properties of CaFeAsH will help to bridge the gap between parent 1111-type and other 11-, 111-, and 122-type iron-based superconductors.

Acknowledgement

We thank Prof. Emer. O. Fukunaga, Assoc. Prof. H. Mizoguchi, Drs. T. Atou, T. Hanna, and J. Bang for valuable discussions. This study was supported by MEXT Element Strategy Initiative to form Core Research Center. H. Hi was also supported by Support for Tokyotech Advanced Research (STAR). The neutron scattering experiment was approved by the Neutron Scattering Program Advisory Committee of IMSS, KEK (Proposal No. 2014S06).

Figure and Table captions

**Figure 1: (a)** Temperature dependence of specific heat $C_p$ of CaFeAsH. The solid line denotes a fitting curve calculated by Eq. (2)-(4). The $C_p$ data from 90 to 105 K were excluded for this fitting. **(b)** The $C_p$ change ($\Delta C_p/T$) and entropy change ($\Delta S$) through the phase transition. Red and blue symbols represent $\Delta C_p/T$ and $\Delta S$, respectively. **(c)** A schematic time relaxation curve of the sample temperature crossing the first-order transition point. A plateau indicated by arrows appears in the relaxation curve. **(d)** Time relaxation curves crossing the transition temperature. Solid lines are fitting curves calculated from the linear combination of two exponential functions with different time constants. **(e)** Comparison of $C_p$ close to the phase transition temperature obtained by different relaxation methods. Red and black symbols show $C_p$ data obtained by normal relaxation and scanning methods, respectively.

**Figure 2: (a)** Temperature dependence of XRD peak profile reflected from tetragonal 221 plane. **(b)** Temperature dependence of the structural order parameter, $P = (a_O - b_O)/(a_O + b_O) \times 10^3$. **(c)** Temperature dependence of the lattice constants of CaFeAsH. The lattice parameters $a$ and $b$ in the tetragonal phase above $T_s$ ($a_T$ and $b_T$) is multiplied by $\sqrt{2}$.

**Figure 3: (a) and (b)** Neutron powder diffraction patterns of CaFeAsD at (a) 150 K and

(b) 10 K. Black crosses and the red line represent observed and calculated intensities, respectively. The range from 4.7 to 4.8 Å was excluded from the Rietveld analysis owing to the presence of an unidentified peak. **(c)** Crystal structure of low temperature orthorhombic CaFeAsD. The unit cell was expressed as a $\sqrt{2}a \times \sqrt{2}a \times 2c$ magnetic supercell containing eight chemical formula. **(d)** AFM structure of CaFeAsD. The spins on Fe atoms are shown.

**Figure 4:** Temperature dependence of the electrical resistivity $\rho$ (open circles) and its temperature derivative $d\rho/dT$ (filled circles) of CaFeAsH.

**Figure 5:** Plot of $T_s - T_N$ of various parent iron-based superconductors versus FeAs interlayer distance. Red, green, blue circles show data of CaH-, CaF-, and SrF-1111, respectively.[68,69] Red, orange, yellow, green, light blue, blue, and purple squares are La-, Ce-, Pr-, Nd-, Sm-, Gd-, and Tb-1111, respectively.[61,63,79–82] Black, red, green, and blue triangles are Ca-, Eu-, Sr-, and Ba-122.[25,83–85] Black crosses and stars are NaFeAs and FeTe, respectively.[86,87] The deviation of NaFeAs from the general tendency is attributed to doped carriers produced by a partial Na deficiency.[87]

**Table 1**: Structural parameters and physical properties of CaFeAsH and CaFeAsF.

Figures and Tables

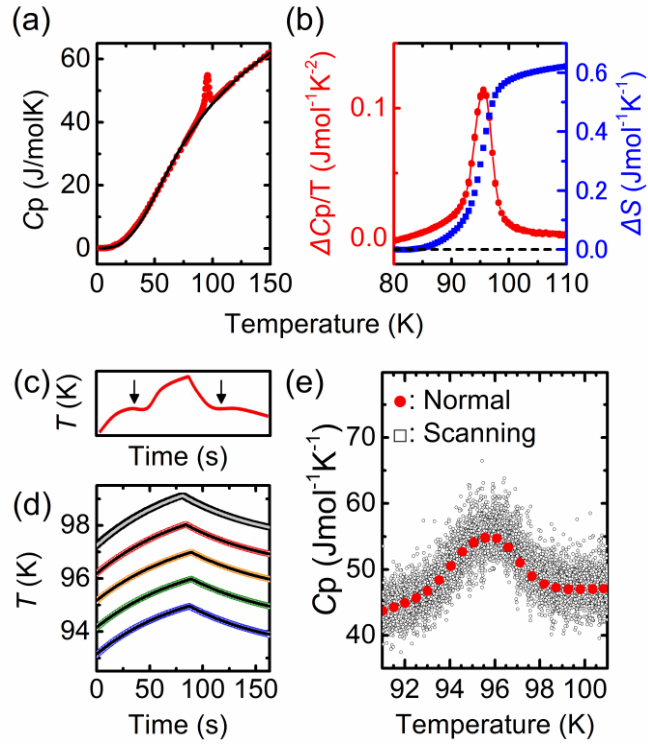

Figure 1

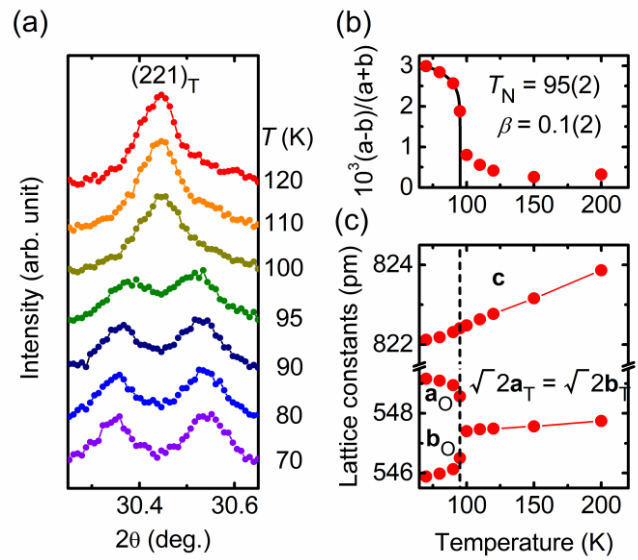

Figure 2

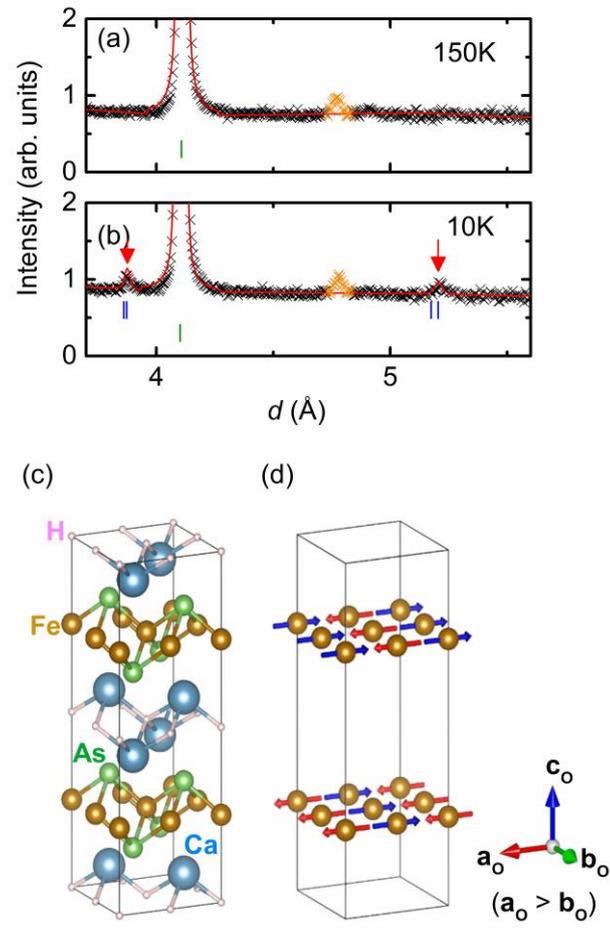

Figure 3

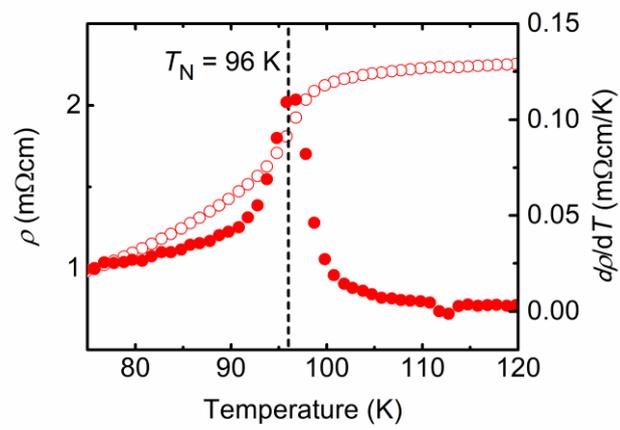

Figure 4

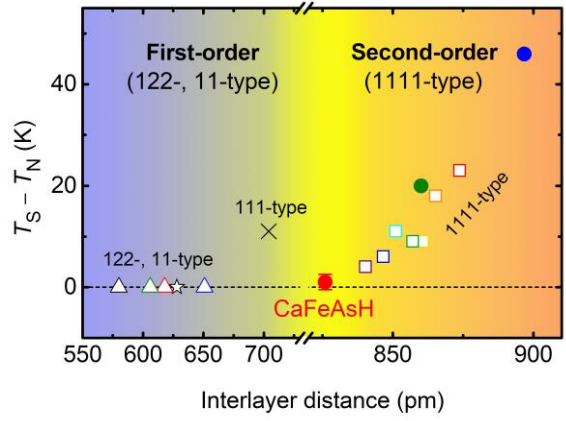

Figure 5

Table 1

|  | CaFeAsH | CaFeAsF |
|---|---|---|
| $a$ (pm) | 387.87 [a] | 387.84 [b] |
| $c$ (pm) | 826.02 [a] | 859.26 [b] |
| $h_{As}$ (pm) | 141.61 [a] | 140.47 [b] |
| $h_{Ca}$ (pm) | 123.75 [a] | 130.61 [b] |
| $d_{Ca-As}$ (pm) | 147.41 [a] | 157.99 [b] |
| $Bm$ (GPa) | 61.28 [c] | 82.79 [c] |
| $T_s$ (K) | 95 (2) | 134 (3) [d] |
| $T_N$ (K) | 96 | 114 (3) [d] |

a reference 19

b reference 20

c reference 36

d reference 61